\documentclass[sigconf]{aamas} 

\usepackage{balance} 
\usepackage[utf8]{inputenc} 
\usepackage[T1]{fontenc}    
\usepackage{hyperref}       
\usepackage{url}            
\usepackage{booktabs}       
\usepackage{amsfonts}       
\usepackage{nicefrac}       
\usepackage{microtype}      
\usepackage{xcolor}         
\usepackage{soul}
\usepackage{comment}
\usepackage{bbm}

\usepackage{amssymb,amsmath,mathrsfs}
\usepackage{comment,multirow,graphicx,bbm,subcaption,caption,multirow,mathtools}
\usepackage{algorithmic,algorithm}
\usepackage{xcolor}
\usepackage{wrapfig}
\usepackage{balance}

\setcopyright{ifaamas}
\acmConference[AAMAS '22]{Proc.\@ of the 21st International Conference
on Autonomous Agents and Multiagent Systems (AAMAS 2022)}{May 9--13, 2022}
{Online}{P.~Faliszewski, V.~Mascardi, C.~Pelachaud,
M.E.~Taylor (eds.)}
\copyrightyear{2022}
\acmYear{2022}
\acmDOI{}
\acmPrice{}
\acmISBN{}

\acmSubmissionID{153}

\title[Q-Learning in Population Games]{The Dynamics of Q-learning in Population Games:\\ a Physics-Inspired Continuity Equation Model}

\author{Shuyue Hu$^\dagger$, Chin-Wing Leung$^\ddagger$, Ho-fung Leung$^\ddagger$, Harold Soh$^\dagger$}
\affiliation{
\institution{National University of Singapore$^\dagger$, The Chinese University of Hong Kong$^\ddagger$}
\country{Singapore$^\dagger$, Hong Kong SAR, China$^\ddagger$}}
\email{shuyuehu217@gmail.com, {cwleung, lhf}@cse.cuhk.edu.hk,  harold@comp.nus.edu.sg }

\begin{abstract}
Although learning has found wide application in multi-agent systems, its effects on the temporal evolution of a system are far from understood. This paper focuses on the dynamics of Q-learning in large-scale multi-agent systems modeled as population games.
We revisit the replicator equation model for Q-learning dynamics and observe that this model is inappropriate for our concerned setting. Motivated by this, we develop a new formal model, which bears a formal connection with the continuity equation in physics. 
We show that our model always accurately describes the Q-learning dynamics in population games across different initial settings of MASs and game configurations.
We also show that our model can be applied to different exploration mechanisms, describe the mean dynamics, and be extended to Q-learning in 2-player and n-player games.
Last but not least, we show that our model can provide insights into algorithm parameters and facilitate parameter tuning.
\end{abstract}


\newcommand{\BibTeX}{\rm B\kern-.05em{\sc i\kern-.025em b}\kern-.08em\TeX}

\begin{document}


\pagestyle{fancy}
\fancyhead{}


\maketitle

\section{Introduction}
Recent years have witnessed a significant gain in the learning capability of intelligent agents.
These advances have spurred the usage of learning agents in many \emph{large-scale} multi-agent systems (MASs) that are concerned with a great number of agents, such as autonomous vehicles for transportation \citep{sallab2017deep}, online trading/bidding agents in financial markets~\citep{wang2016display}, and cooperative robots for search and rescue~\citep{long2018towards}.
However, despite wide application, learning in large-scale MASs is far from understood and its theoretical underpinnings remain elusive.

\begin{figure}
     \centering
     \begin{subfigure}[t]{0.9\columnwidth}
         \centering
         \includegraphics[height=1.42in]{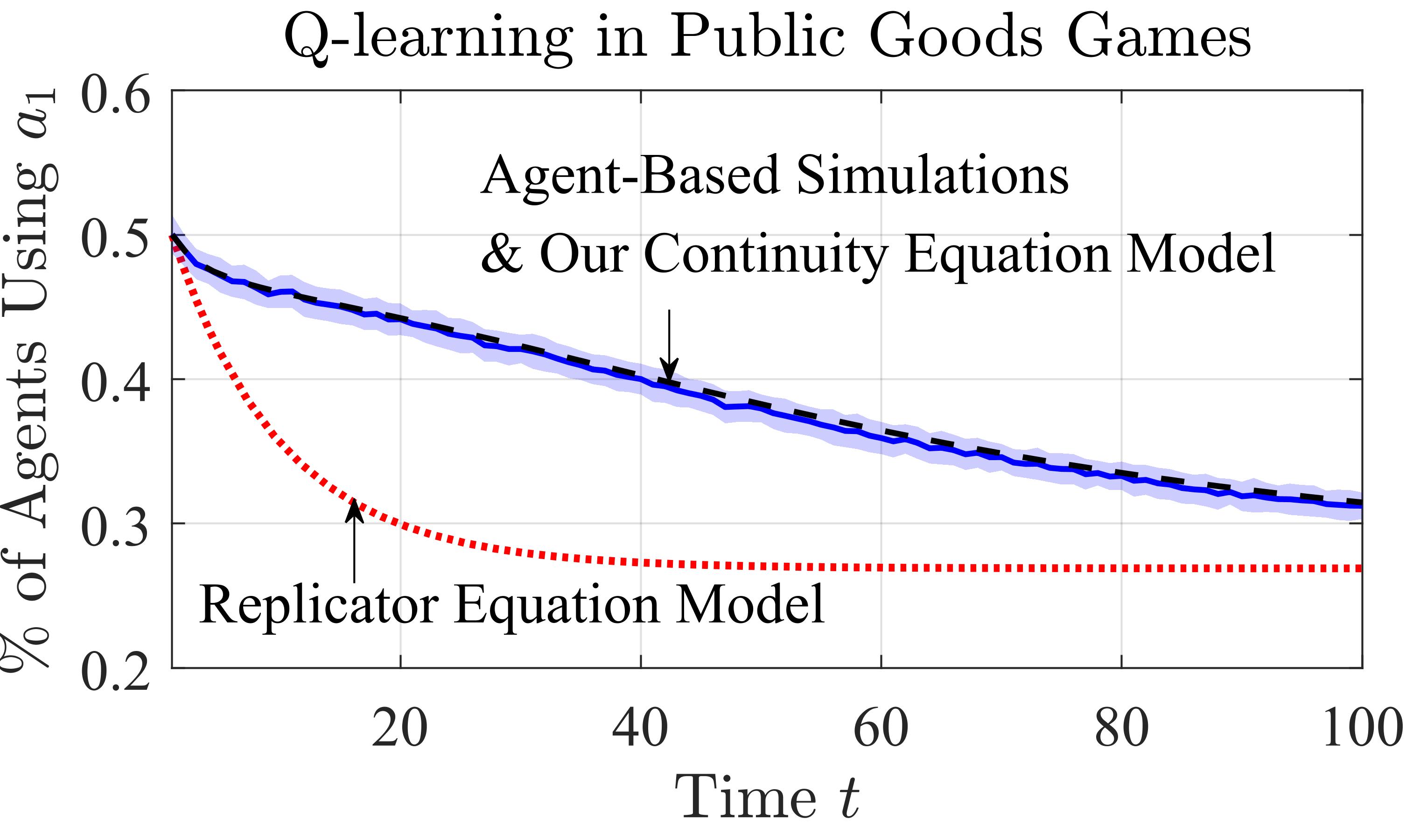}
         \caption{\small Homogeneous MAS. Agents have the same initial Q-values which are both 4 for two actions $a_1$ and $a_2$. } 
     \end{subfigure}
     \hfill
     \begin{subfigure}[t]{0.9\columnwidth}
         \centering
         \includegraphics[height=1.42in]{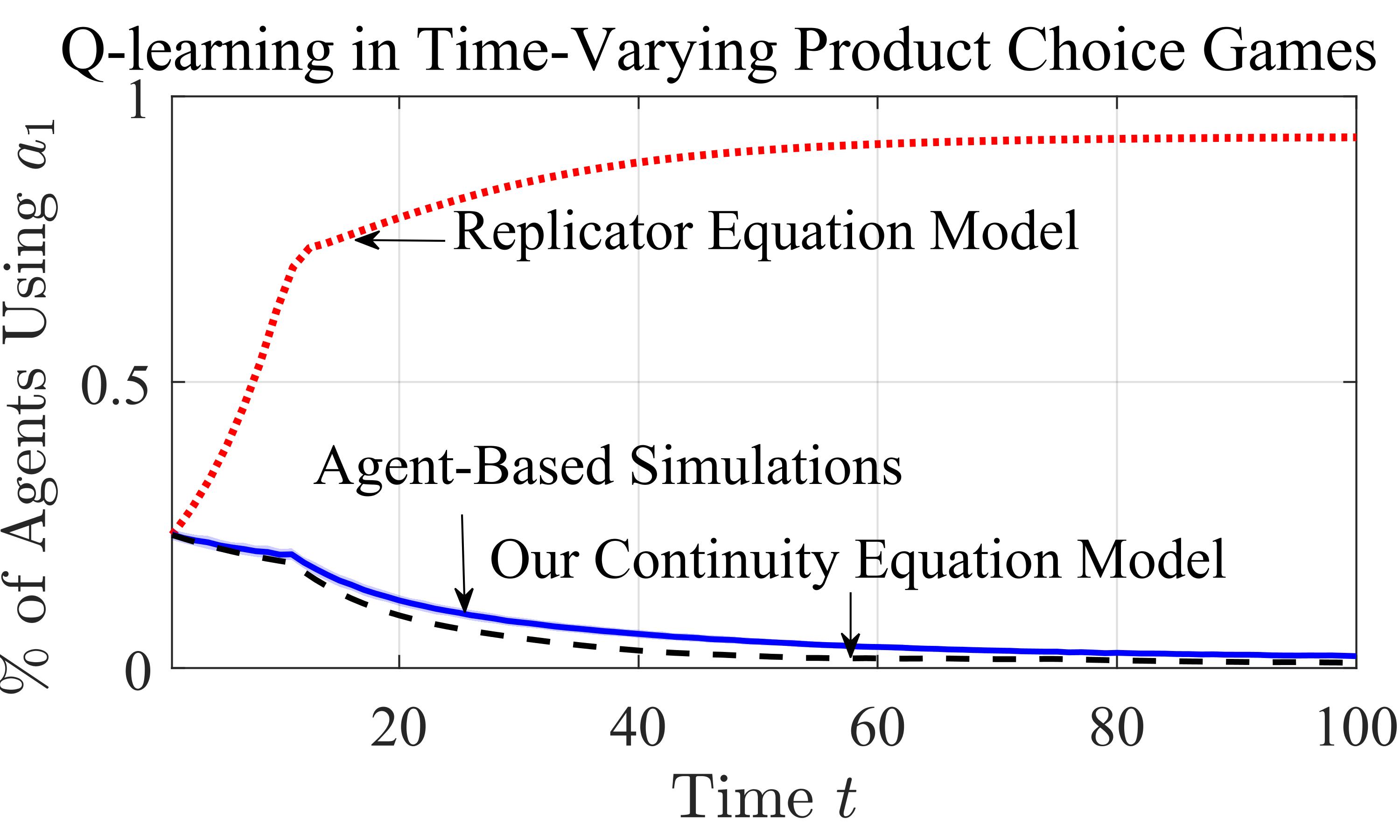}
         \caption{ \small Heterogeneous MAS. 
          Initial Q-values for  actions $a_1$ and $a_2$ are distributed according to $ \mathrm{Beta}(15,30)$ and $  \mathrm{Beta}(10,10)$, respectively, with support $[-1.5,1.5]$.  }
     \end{subfigure}
     \vspace{-0.1in}
     \caption{\small 
     Comparison among the population dynamics described by the Replicator Equation Model~\cite{tuyls2003selection,sato2003coupled} (REM, dotted red line) and our Continuity Equation Model (CEM, dashed black line), and the actual dynamics averaged over 100 runs of agent-based simulations (shaded blue line with the shaded area representing the standard deviation). The game configurations are summarized in Table 1, 
     the Boltzmann temperature is $3$, and the learning rate is $0.1$. In both homogeneous and heterogeneous MASs, our CEM better captures the qualitative and quantitative dynamics of the systems.}
     \vspace{-0.1in}
\end{figure}

Population games are canonical models of strategic interactions of large-scale MASs~\cite{sandholm2010population}.
Traditionally, a multi-agent learning (MAL) algorithm is often examined by whether the strategy profile will converge to a (e.g., Nash) equilibrium in games~~\citep[e.g.,][]{fudenberg1998theory,subramanian2019meanfield,yang2018mean}.
However, emergent theoretical research has shifted its focus to  the \emph{dynamics} because static equilibrium notions are fundamentally limiting --- they cannot express any temporal evolution of a system nor long-term non-equilibrium phenomena~\citep[e.g.,][]{bailey2019multi,Cheung2018,vlatakis2019poincare,sato2002chaos}.
As Tuyls and Parsons~\citep{tuyls2007evolutionary} voice, the development of theory in this direction is crucial because it will not only yield a better theoretical understanding of existing algorithms, but potentially facilitate
the design of new methods, leading to practical algorithmic advancements.

In this work, we focus on the dynamics of Q-learning in population games.
Q-learning, as proposed by Watkins \& Dayan~\cite{watkins1992q}, is one of the most important learning algorithms in AI literature. 
It forms the basis of numerous learning methods and is a main focus of many theories in MAL~\citep[e.g.,][]{Gomes2009dynamic,wunder2010classes,kianercy2012dynamics}. 
In their seminal works, Tuyls et al.~\cite{tuyls2003selection} and Sato \& Crutchfield~\cite{sato2003coupled} proposed the \emph{replicator equation model} (REM)\footnote{In~\cite{tuyls2003selection}, the model is called the selection-mutation model.} to describe the dynamics of agents that apply Q-learning with Boltzmann exploration in 2-player normal-form games. 
The REM reveals a surprising connection between multi-agent Q-learning and the well-known \emph{replicator dynamics} of  evolutionary game theory (EGT). This connection paved the way to the study Q-learning from the EGT perspective and has inspired many works in the MAL literature~\citep[e.g.,][]{gatti2013efficient,hennes2020neural,galstyan2013continuous}.
More recently, Leonardos et al. applied the REM to n-player games~\cite{leonardos2021exploration} and population games with homogeneous populations~\cite{leonardos2020catastrophe}; using the REM as an example, they provided new mechanisms to induce phase transitions between multiple equilibria in MASs.

Although many studies of Q-learning dynamics are based on the REM~\citep[e.g.,][]{kianercy2012dynamics,panozzo2014evolutionary,hennes2009state}, we observe that the REM is \emph{inappropriate} for Q-learning in general population games. To elaborate, the REM was designed for multi-player games with a discrete number of agents and simplifies the canonical Q-learning dynamics by (i) tracking only the policies of individual agents,
and (ii) assuming each agent is  performing multiple updates of the Q-values for each update of the policy. While these simplifications are natural in certain settings, they cause the model to neither (i) differentiate between agents that have different Q-values but happen to have the same policy at a given time step, nor (ii) capture the effects of the asynchronous update in Q-learning. As such, the REM can be inexact when applied to Q-learning in population games which feature large and generally heterogeneous populations. 
As shown by the example in Figure 1, the dynamics prescribed by the REM do not match the actual dynamics in agent-based simulations; sometimes, the REM even suggests a system outcome that is completely different than the ground truth (Figure 1(b)).

Motivated by this observation,
we develop a new formal model for Q-learning in population games.
Rather than only tracking agent policies, we directly track the Q-values of individual agents.
Moreover, we propose to tackle the asynchronous update of Q-learning by modeling its stochastic effects on Q-values.
Note that unlike 2-player games, population games involve infinitely many agents that typically have diverse initial Q-values and develop different policies afterwards. This poses a new challenge: 
how can we characterize the effects of population heterogeneity on Q-learning dynamics? 
To address this challenge,
we focus on the distribution of Q-values in the population; in particular, we investigate the evolution of this distribution function as time progresses, and derive a \emph{differential equation} to model its temporal evolution.
Our proposed solution is inspired by statistical physics, where studying the dynamics of a probability distribution rather than the dynamics of individuals is a classic approach (examples range from the heat transfer equation to the Fokker–Planck equation for Brownian motion).

The resultant model (Equation~\ref{eq:pde}) from our  approach takes the form of a partial differential equation (PDE), which is fundamentally different from the REM that is based on ordinary differential equations (ODEs).
In particular, we find that our model can be viewed as a \emph{continuity equation} that describes the transport phenomena (e.g., of mass or energy) in a physical system.
This suggests a connection between MAL and physics  --- the Q-learning dynamics in population games is analogously the transport of the agent mass in the Q-value space. 
Moreover, we observe that our continuity equation model (CEM) has some interesting properties (Section 5.2) --- CEM can (i)
be applied to different exploration mechanisms, (ii) describe the mean dynamics~\cite{sandholm2015population} (the temporal evolution of the mean policy) in the system,
(iii) be reduced to a system of coupled ODEs for homogeneous populations, 
and (iv) be extended to model Q-learning dynamics in 2-player games and n-player games. 

In our experiments, we validate that given different population games and initial settings of MASs, our CEM always provides an accurate description of Q-learning dynamics with respect to the actual dynamics in agent-based simulations (Section 6.1).
In addition, we illustrate two potential use cases of our model. 
Through a concrete example, we show that our CEM can provide non-trivial insights into the effects of algorithm parameters (particularly, the temperature of Boltzmann exploration); these insights lead to practical guidelines for notoriously cumbersome parameter tuning (Section 6.2).
We also show that our CEM can contrast the dynamics that arise from different exploration mechanisms, which potentially can facilitate the choice of exploration mechanisms (Section 4 in the supplementary). The supplementary of this paper  can be found online~\cite{hugooglesites,clearsite}.

To summarize, our key contributions are:
\vspace{-0.05in}
\begin{itemize}
    \item An analysis of the limitations of the well-known replication equation model in settings with heterogeneous agents that perform asynchronous updates;
    \item The development of a new theoretical model for Q-learning dynamics in population games, which bears a formal connection with the continuity equation in physics;
    \item Experimental results validating the descriptive power of our continuity equation model, and two examples illustrating its use to gain insights into the effects of algorithm parameters and to contrast the dynamics that arise from different exploration mechanisms. 
\end{itemize}

\section{Related Work} 
Previous works that examine Q-learning dynamics are largely based on the REM proposed by 
Tuyls et al.~\cite{tuyls2003selection} and Sato \& Crutchfield~\cite{sato2003coupled}.
Panozzo et al.~\cite{panozzo2014evolutionary} introduced an extension of the REM for Q-learning that operates on sequence forms.
Based on the REM, Kianercy \& Galstyan~\cite{kianercy2012dynamics} provided a comprehensive characterization of the fixed point structure for Q-learning in different 2-player games.
Kaisers \& Tuyls~\cite{kaisers2012common} noticed that the prediction of the REM in 2-player games may deviate from the actual Q-learning dynamics;  
but rather than developing a more accurate model for Q-learning,
they proposed a new algorithm that is more consistent with what the REM predicts.
More recently, the REM has been applied to study phase transitions in 
$n$-player games~\cite{leonardos2021exploration} and population games where agents have the same initial policy~\cite{leonardos2020catastrophe};
Leonardo et al.~\cite{leonardos2021exploration,leonardos2020catastrophe} showed that by tuning the exploration parameter, there are phase transitions between multiple equilibria in MASs.
The REM has inspired many works to study MAL (not limited to Q-learning) using EGT approaches; we refer interested readers to a recent survey~\cite{bloembergen2015evolutionary} and references therein.
However, the REM is unable to provide an appropriate model of Q-learning in general population games due to the simplifications it makes.

There are few exceptions that examine Q-learning dynamics without the use of REM and its variants~\cite{Gomes2009dynamic,wunder2010classes,hu2019modelling}. 
Gomes \& Kowalczyk~\cite{Gomes2009dynamic} and Wunder et al.~\cite{wunder2010classes} 
focused on Q-learning with $\epsilon$-greedy exploration in 2-player games; however, their approaches are tailored to address the discontinuity caused by $\epsilon$-greedy exploration and are not applicable to large agent populations.
Hu et al.~\cite{hu2019modelling} considered an  $n$-agent setting where Q-learning agents with Boltzmann exploration are paired up to play 2-player games; using mean field theory, they reduced the setting to an 2-agent setting and developed a Fokker-Planck equation for the learning dynamics.
As we shall discuss in Section 5.2, 
our model can be generalized to their setting, even though 2-player games and population games are different in nature.
In this sense, our model can be viewed as a generalization of ~\cite{hu2019modelling}, which goes beyond 2-player games and Boltzmann exploration.

Lahkar \& Seymour~\cite{lahkar2013reinforcement} studied Cross learning in population games and also derived a continuity equation for the learning dynamics. 
In addition to the difference in algorithms (Cross learning is policy-based whereas Q-learning is value-based), their approach is \emph{incompatible} with Q-learning.
Specifically, their approach is to work with the distribution of the policy in the population.
However, as we shall show in Section 4.2, tracking the policies of agents can be misleading for Q-learning (especially in population games); the policy dynamics are not equivalent to the Q-values dynamics.

Mean field games are also concerned with infinitely many agents \cite{guo2019learning,subramanian2019meanfield}. 
Lasry \& Lions~\cite{lasry2007mean} developed a system of two coupled PDEs --- a Fokker-Planck equation and a Hamilton–Jacobi–Bellman equation --- for the mean-field game theory setting.
However, rather than Q-learning, agents in~\cite{lasry2007mean} apply optimal control to a well understood system with complete observation of the system state.

\section{Preliminaries}
In this paper, we consider Q-learning in population games. 
Specifically, at each time step $t$, a population of  Q-learning agents each takes an action independently. Based on the action applied and the population state, each agent receives an immediate reward in the game and adapts its Q-value and policy accordingly.
At the next time step $t+1$, agents start over for another play of the game.
In this section, we define population games and Q-learning.

\subsection{Population games}
The population game is a widely adopted framework for modeling strategic interactions that are commonly observed in large-scale MASs~\citep{sandholm2015population}, such as network congestion, task allocation, and social norm emergence.
Specifically, population games model scenarios that simultaneously exhibit three properties:
(i) the number of agents is large, (ii) each agent is small, such that any particular one agent’s behavior has little or negligible effect on other individual agents, and (iii) each agent is anonymous, in that exchanging the labels of agents will not create any difference. 

Consider a set $\mathcal{N}=\{1, \ldots, n\}$ of $n$ agents with $n\to \infty$ and a set $\mathcal{A}=\{a_1, \ldots ,a_m\}$ of $m$ actions available to each agent.
Suppose that a population game will be played for $T$ time steps.
An agent's payoffs in a population game depend only on its own behavior and the aggregated effect of the other agents' behaviors which is usually termed as \emph{population state}.
For every time step $t$, the population state is represented by a vector $\vec{o}_t=[o_{1,t}, \ldots,\ o_{m,t}]^\top$, where $o_{j,t}$ is the proportion of agents taking action $a_j \in\mathcal{A}$ in the population at time $t$.
The reward function is given by 
$R(a_j, \vec{o}_t,t)$
which determines the payoff of an agent by the action $a_j$ it uses and the population state $\vec{o}_t$ at time $t$. 
In general, the reward function of a population game can change over time.
The population state evolves as agents interact with one another. 

\subsection{Q-learning and Boltzmann Exploration}  
Q-learning~\cite{watkins1992q} is typically defined in the context of a Markov decision process (MDP). 
In this work, we focus on population games where there are no environmental state transitions. Environmental statelessness is a common assumption made in theory for MAL~\citep[e.g.,][]{bailey2019multi,Cheung2018,hu2019modelling} and simplifies analysis and exposition.
A stateless MDP consists of a set $\mathcal{A}$ of available actions and an immediate reward function that gives the reward of using each action.
For a stateless MDP, Q-learning maintains a $Q$-value for each action.
Consider an arbitrary Q-learning agent $i$.
We define the set of Q-values of agent $i$ at time $t$ to be $\vec{Q}_{t}^{i}=[Q_{1,t}^{i}, \ldots, Q_{m,t}^{i}]^\top$ where $Q_{j,t}^{i}$ is the Q-value for action $a_j$.
Suppose that at time $t$, agent $i$ plays the action $a_j$ and receives an immediate reward $r_{j,t}^{i}=R(a_j, \vec{o}_t, t)$ determined by the reward function of the population game. 
The agent $i$ will update the $Q$-value of action $a_j$ as follows:
\begin{equation}
\label{eq:q}
Q_{j, t+1}^{i}= (1-\alpha)  Q_{j, t}^{i} +\alpha r_{j,t}^{i}
\end{equation}
where $\alpha$ is the learning rate. Note that for every time step, only the Q-value of the action in use is updated; the Q-values of the other actions (that are not applied at this time step) remain \textit{unchanged}.

There are multiple mechanisms for a Q-learning agent to select an action based on its Q-values. 
We define the policy of agent $i$ at time $t$ to be $\vec{x}_{t}^{i}=[x_{1,t}^{i} , \ldots, x_{m,t}^{i}]^\top$ where  $x_{j,t}^{i}$ is the  probability that agent $i$ uses action $a_j$. 
For Boltzmann exploration,
the value of  $x_{j,t}^{i}$  is given by $x_{j,t}^{i}={e^{\tau Q_{j,t}^{i}}}/ [\sum_{k=1}^{m}{e^{\tau Q_{k,t}^{i}}}  ]$,
where $\tau \in [0,\infty)$ is the Boltzmann temperature that controls how much the agent explores.
The agent is in pure exploration (randomly taking each action) when $\tau$ is $0$, and in pure exploitation (greedily taking the action with the highest $Q$-value) when $\tau \to \infty$.

\section{Replicator Equation Model Revisited}
In this section, we revisit the REM~\cite{tuyls2003selection,sato2003coupled} for Q-learning with Boltzmann exploration.
Tuyls et al.~\cite{tuyls2003selection} and Sato \& Crutchfield~\cite{sato2003coupled} developed this model for Q-learning in 2-player games. 
Recent work~\cite{leonardos2020catastrophe} has applied the REM to population games with homogeneous populations. 
We describe this model in Section 4.1. 
In Section 4.2, we analyze two simplifications that this model makes for Q-learning.
In Section 4.3, we discuss the application of this model to population games and show that this model can provide inexact predictions under this setting.
For ease of presentation, in this paper, we consider a set $\mathcal{A}=\{a_1, a_2\}$ of two actions;
generalization of our analysis/approach to cases with more than two actions is straightforward.

\subsection{Replicator Equation Model}
In their seminal work, Tuyls et al.~\cite{tuyls2003selection} and Sato \& Crutchfield~\cite{sato2003coupled}  developed replicator equations to model the dynamics of Q-learning with Boltzmann exploration in 2-player games.
Let $i$ denote an arbitrary player in a 2-player game.
The replicator equation that models the time evolution of the policy of agent $i$ is given as follows:

\begin{equation}
\label{eq:replicator}
    \frac{dx_{j,t}^{i}}{dt} = \alpha \tau  \underbrace{x_{j,t}^{i} \left(r_{j,t}^{i} - \sum_{a_k\in \mathcal{A}}x_{k,t}^{i} r_{k,t}^{i} \right) }_{T_1} + \alpha  \underbrace{ x_{j,t}^{i} \sum_{a_k \in \mathcal{A}} x_{k,t}^{i} \ln \frac{x_{k,t}^{i}}{x_{j,t}^{i}}}_{T_2}
\end{equation}
where $x_{j,t}^{i}$ is the probability that agent $i$ uses any action $a_j \in \mathcal{A}$ at time $t$.
Note that the term $T_1$ is exactly the well-known replicator dynamics capturing the selection mechanism in EGT, and the term $T_2$ can be decomposed into two entropy terms handling the mutation mechanism in EGT~\cite{tuyls2003selection}.
Therefore, this model elegantly brings forward the connection between multi-agent Q-learning and EGT.

\subsection{\textcolor{black}{Replicator Equation Model vs Q-learning}}
\label{sec:remvsq}

\paragraph{Representation of Q-learners.}
One simplification in the REM is that 
it represents agents with their policies and does not differentiate between agents that have different Q-values but the same policy.
Suppose at time $t$, agents $i$ and $j$ have the same policy but different Q-values, and both apply action $a_1$.
Equation \ref{eq:replicator} dictates that if two agents have the same learning parameters and reward functions, they will develop \emph{exactly the same} policy; the changes in their polices do not explicitly depend on their Q-values.
However, this is generally \emph{not} true. Consider the policy of agent $i$ for time $t+1$ 
\begin{equation}
\begin{aligned}
        x_{1,t+1}^{i} & =\frac{e^{\tau Q_{1,t+1}^{i}}} {e^{\tau Q_{1,t+1}^{i}} + e^{ \tau Q_{2,t+1}^{i}}} = \frac{1}{ 1 + e^{\tau \left( Q_{1,t+1}^{i} -  Q_{2,t+1}^{i} \right)} } \\
         &  =  \frac{1}{ 1 + e^{\tau \left[  Q_{1,t}^{i} - Q_{2,t}^{i} + \alpha r_{1,t}^{i} - \alpha Q_{1,t}^{i} \right ] } }
\end{aligned}
\end{equation}
and $x_{2,t+1}^{i} = 1 - x_{1,t+1}^{i}.$ 
The precondition of agents $i$ and $j$ having the same policy at time $t$ only ensures $Q_{1,t}^{i} - Q_{2,t}^{i} = Q_{1,t}^{j} - Q_{2,t}^{j}$ (this can be inferred from the second equality).
Therefore, 
agents will not necessarily develop the same policy for time $t+1$ if they do not have the same Q-values at time $t$.
This observation suggests that 
representing a Q-learner with its policy and  tracking only its policy may not provide a good description of its dynamics. 

\paragraph{Update frequency of Q-values.} Another difference between the REM and  Q-learning is that the REM implicitly assumes that at each time step, a Q-learner will update the Q-values for \emph{every} action rather than only the action in use.
To see this, we make use of the equality $\ln ({x_{2,t}^ i}/{ x_{1,t}^i} ) 
= \tau (Q_{2,t}^{i} - Q_{1,t}^{i} ) $, and rewrite Equation 1 as
\begin{equation}
\label{eq:dxdt}
\begin{aligned}
\frac{dx_{1,t}^{i}}{dt}  
= \frac{\partial x_{1,t}^{i}}{\partial  Q_{1,t}^{i}} \alpha (r_{1,t}^{i} - Q_{1,t}^{i} ) + \frac{\partial x_{1,t}^{i}}{\partial  Q_{2,t}^{i}} \alpha (r_{2,t}^{i} - Q_{2,t}^{i} ) .
\end{aligned}
\end{equation}
By the chain rule, we also have $\frac{dx_{1,t}^{i}}{dt} =\frac{\partial x_{1,t}^{i}}{\partial  Q_{1,t}^{i}}  \frac{d Q_{1,t}^{i}}{dt} +  \frac{\partial x_{1,t}^{i}}{\partial  Q_{2,t}^{i}}  \frac{d Q_{2,t}^{i}}{dt}$, which suggests 
the model assumes that for each action $a_j \in \mathcal{A}$,
   $\frac{d Q_{j,t}^{i}}{dt} = \alpha (r_{j,t}^{i} - Q_{j,t}^{i} ).$ 
From this, we see that the Q-value of \emph{every} action is always updated at a given time step. 
This contradicts the standard asynchronous update rule of Q-learning --- only the action in use should be updated.

Note that our above analysis is \emph{not} limited to only a specific type of games (e.g. 2-player games). 
In other words, the above two issues in representation and update frequency are inherent in the model no matter what games the model is applied to.

Importantly, we emphasize that we do \emph{not} claim that the REM is wrong or inferior in general.
Indeed, if one considers that each agent updates its Q-values for all actions synchronously,  the simplifications pointed out above will vanish, and the REM will provide an accurate and precise description of the learning dynamics. 
There are two possibility for such synchronous updates: (i) agents perform many interactions before updating their Q-values (or put differently, the learning dynamics is very slow compared
to interactions~\cite{sato2003coupled}), and (ii) agents apply the Frequency Adjusted Q-learning~\cite{kaisers2010frequency}.\footnote{Kaisers and Tuyls~\cite{kaisers2010frequency} reported a similar finding on the update frequency that the model assumes. They argued that the behaviors predicted by the REM are more desirable and proposed the Frequency Adjusted Q-learning whose dynamics in 2-player games is more consistent with what the REM predicts.}
Nevertheless, as defined by Watkins \& Dayan~\cite{watkins1992q}, the \emph{asynchronous} update rule of Q-learning is standard and important; this is a norm in the literature for Q-learning and its variants (examples include impactful algorithms~\cite{hu2003nash,hasselt2010double,sutton2018reinforcementsarsa}). 
Therefore, the simplifications pointed out above are  non-trivial and require formal treatment for the dynamics of Q-learning.

\subsection{Application to Population Games}

Recent work~\cite{leonardos2020catastrophe} has applied the REM to population games in \textit{homogeneous} MASs where all agents have the same initial policy.
Due to the symmetry of agents, the superscript $i$ in Equation~\ref{eq:replicator} can be dropped; thus, the model  describes how the policy, which is the same for every agent, evolves as time $t$ progresses.
This model can also be applied to population games in \textit{heterogeneous} MASs where agents have diverse initial Q-values and start with different policies.
To achieve this, in Equation~\ref{eq:replicator}, one can replace $x_{j,t}^{i}$  with $o_{j,t}$ and  $r_{j,t}^{i}$ with $R(a_j, \vec{o}_t ,t)$;
here, Equation~\ref{eq:replicator} models the dynamics of population action frequencies.\footnote{Alternatively, one can maintain an separate Equation~\ref{eq:replicator} for each initial policy, but this approach is intractable due to infinitely many agents.}

We hypothesize that the two issues pointed out in Sec. \ref{sec:remvsq} above, coupled with potential population heterogeneity in population games, conspire to cause inexact descriptions under the concerned setting.
Intuitively, because the REM implicitly assumes synchronous updates of Q-values, the learning speed predicted by the model is likely to deviate.
In addition, because the REM considers agents that have the same policy but different Q-values to be identical, the effects of population heterogeneity are underestimated. 
Unlike 2-player games, population heterogeneity generally plays an important role in population games given infinitely many agents.
Thus, the REM may provide a less accurate description of Q-learning dynamics in population games than in 2-player games.

To verify our hypothesis, we compare the dynamics predicted by  the  model against agent-based simulation results (which are the ground truth), given the same initial settings of MASs. 
In this work, for each comparison, we performed 100 independent simulation runs to  generate the simulation results; for each run, there were 1,000 agents. 
It is clear in Figure 1(a) that there is a noticeable discrepancy in the speed of convergence even for homogeneous MASs playing the relatively simple public goods game (where there is a unique Nash equilibrium).
As shown in Figure 1(b), for heterogeneous MASs playing the time-varying product choice game (where there are two pure-strategy Nash equilibria), the model predicts a system outcome that is completely different from the ground truth.
To be more specific, the model predicts that the population would quickly flock to use action $a_1$; however, in 100 simulation runs, the population always converged to use action $a_2$.

In summary, when applied to population games, the REM can provide inexact predictions on both \emph{speeds} and \emph{outcomes} of Q-learning.
Hence, we caution against using this model when examining Q-learning in population games.

\section{Continuity Equation Model}

In this section, we present a new model --- the continuity equation model (CEM, Equation \ref{eq:pde}) --- which provides an accurate description of Q-learning in population games. The two issues of the REM and the potential heterogeneity of MASs are non-trivial to address.
Our approach is inspired by statistical physics, where studying the dynamics of a probability distribution rather than the dynamics of individuals, greatly reduces the degrees-of-freedom involved and simplifies the analysis.
In Section 5.1, we highlight the key steps in the development of our model, and 
in Section 5.2, we discuss its key properties.

\subsection{Development of the Model}

The key idea underlying our approach is to work with the distribution of Q-values in the population and derive a differential equation that describes the temporal evolution of this distribution.
By working with the distribution of Q-values, we represent agents with their Q-values and the issue caused by representing Q-learners with their policies disappears.
In addition, we address the asynchronous update of Q-values in our model by capturing its stochastic effect on Q-values.

Let $M_n(\vec{q},t)$ where $\vec{q}=[q_1,q_2]^\top \in \mathbb{R}^2$ be the empirical cumulative distribution function (CDF) of the Q-values in the population at time $t$, i.e.
$M_n(\vec{q},t) =  \frac{1}{n} \sum_{i\in \mathcal{N}} \mathbbm{1} (Q_{1,t}^{i} \leq q_1, Q_{2,t}^{i} \leq q_2)$
where $\mathbbm{1}(\cdot)$ is the indicator function. 
With a slight abuse of notation, let $\vec{Q}_t=[Q_{1,t},Q_{2,t}]^\top \in \mathbb{R}^2$ be a pair of random variables denoting the Q-values of an agent that is randomly drawn from the population at time $t$.
We define $f(\vec{q},t)$ as the probability density function (PDF) for $\vec{Q}_t$ such that the corresponding CDF $F(\vec{q},t)$ is the asymptotic distribution of the empirical CDF $M_n(\vec{q},t)$.
That is,
$f(\vec{q},t)=\frac{d F(\vec{q},t)}{d \vec{q}} $ such that $M_n(\vec{q},t) \overset{\mathcal{D}}{\to} F(\vec{q},t).$
     
We are interested in the time evolution of the PDF $f(\vec{q},t)$. Let $\theta(\vec{q})$ be a test function of Q-values and $\delta\in (0,1]$ be the amount of time that passes between two repetitions of the population game.
We compute the quantity $Y$ defined as 
\begin{equation}
\label{eq:quantity}
Y = \frac{ \mathbb{E}[\theta(\vec{Q}_{t+\delta})] - \mathbb{E}[\theta(\vec{Q}_{t})] } {\delta} 
= \int \theta(\vec{q}) \frac{ f(\vec{q},t+\delta) - f(\vec{q},t) } {\delta} d\vec{q}. 
\end{equation}
Intuitively, $Y$ tracks the change of the expected value of $\theta(\vec{Q}_t)$ between two repetitions of the population game, where the PDF $f(\vec{q},t)$ and $f(\vec{q},t+\delta)$ are generally different after the game play at time $t$.

At time $t$, for an arbitrary agent,
let $\vec{Z}_t = [Z_{1,t},Z_{2,t}]^\top \in \{0,1\}^2$ be a pair of random variables indicating the action applied at time $t$ such that $Z_{j,t}=1$ means action $a_j$ is applied and $Z_{j,t}=0$ means action $a_j$ is not applied. Note that the probability of applying each action is determined by an agent's current Q-values and the exploration mechanism it uses. 
We define such probability as $p_{j}(\vec{q})$ for each action $a_j$.  As such, $Z_{j,t}\sim \mathrm{Bernoulli}\left(p_j(\vec{q}) \right)$.
By the update rule of Q-learning, 
\begin{equation}
\label{eq:qrv}
    \vec{Q}_{t+\delta} = \vec{Q}_t + \delta \alpha \vec{Z}_t \cdot  ( \vec{r}_t - \vec{Q}_t ) 
\end{equation}
where $\vec{r}_t = [r_{1,t},r_{2,t}]^\top$ represents the immediate reward of taking each action and is given by the reward function of the population game. 
Let $\beta =\delta \alpha$.
Based on this equation, 
\begin{equation}
\begin{aligned}
\mathbb{E}[\theta(\vec{Q}_{t+\delta})] &=  \mathbb{E}\left[\theta(\vec{Q}_t + \beta \vec{Z}_t \cdot  ( \vec{r}_t - \vec{Q}_t ) ) \right] \\
&= \int f(\vec{q},t)  \sum_{j \in \{1,2\}} p_j(\vec{q}) \theta( \vec{q} + \beta \vec{e}_j \cdot (\vec{r}_t - \vec{q}) ) d\vec{q}
\end{aligned}
\end{equation}
where $\vec{e}_j$ is the unit vector such that $\vec{e}_1=[1,0]^\top$ and $\vec{e}_2=[0,1]^\top$. 
The Taylor series for $\theta( \vec{q} + \beta \vec{e}_j \cdot (\vec{r}_t - \vec{q}) ) $  at $\vec{q}$ is
\begin{equation}
\begin{aligned}
 & \theta(\vec{q})   +  \left [  \beta \vec{e}_j \cdot  ( \vec{r}_t - \vec{q} ) \right] \partial_{q_j} \theta (\vec{q}) + 
 \frac{1}{2} \left [ \beta \vec{e}_j \cdot  ( \vec{r}_t - \vec{q} ) \right ]^2 \partial_{q_j q_j} \theta (\vec{q}) \\
& \quad  +  o(\left[\beta \vec{e}_j \cdot  ( \vec{r}_t - \vec{q} ) \right]^2).
 \end{aligned}
\end{equation}
Rearranging terms, we obtain
\begin{equation}
\begin{aligned}
 & \mathbb{E}[\theta(\vec{Q}_{t+\delta})] 
  =  \int f(\vec{q},t) \theta(\vec{q}) d\vec{q} \\
 &+ \beta \int f(\vec{q},t)  \sum_{j \in \{1,2\}} p_j(\vec{q})  \left [ \vec{e}_j \cdot  ( \vec{r}_t - \vec{q} ) \right]  \partial_{q_j} \theta (\vec{q})  d\vec{q}  \\ 
 &  +  \frac{\beta^2}{2} \int f(\vec{q},t) \sum_{j \in \{1,2\}} p_j(\vec{q})   \left [  \vec{e}_j \cdot  ( \vec{r}_t - \vec{q} )  \right ]^2 \partial_{q_j q_j} \theta (\vec{q}) d\vec{q} \\
 & + \beta^2 \int f(\vec{q},t) \sum_{j \in \{1,2\}} p_j(\vec{q})  o(\left[ \vec{e}_j \cdot  ( \vec{r}_t - \vec{q} ) \right]^2)  d\vec{q}.
\end{aligned}
\end{equation}
The first term on the right hand side equals $\mathbb{E}[\theta(\vec{Q}_{t})]$.
Moving the first term to the left hand side and dividing both sides by $\delta$, we have
the quantity of interest 
\begin{equation}
\begin{aligned}
Y & = \alpha \int f(\vec{q},t) \sum_{j \in \{1,2\}} p_j(\vec{q}) [\vec{e}_j \cdot  ( \vec{r}_t - \vec{q})]  \partial_{q_j}\theta(\vec{q}) d\vec{q}  \\ 
 &  + \frac{\alpha^2\delta}{2} \int f(\vec{q},t)  \sum_{j \in \{1,2\}}  p_j(\vec{q})   \left [  \vec{e}_j \cdot  ( \vec{r}_t - \vec{q} )  \right ]^2 \partial_{q_j q_j} \theta (\vec{q})  d\vec{q} \\
  & +  \alpha^2  \delta \int f(\vec{q},t)  
  \sum_{j \in \{1,2\} }   p_j(\vec{q})  o(\left[ \vec{e}_j \cdot  ( \vec{r}_t - \vec{q} ) \right]^2)  d\vec{q}. 
\end{aligned}
\label{eq:y}
\end{equation}
Taking the limit of $Y$ with $\delta \to 0$ (assuming the continuous time limit), the contribution of the second and third terms on the right hand side vanishes.

On the other hand, according to the definition of $Y$,
\begin{equation}
\label{eq:y1}
\begin{aligned}
\lim_{\delta \to 0} Y &= \lim_{\delta \to 0} \int \theta(\vec{q}) \frac{ f(\vec{q},t+\delta) - f(\vec{q},t) } {\delta} d\vec{q}  \\
&= \int \theta(\vec{q}) \partial_t  f(\vec{q},t)   d\vec{q}.
\end{aligned}
\end{equation}
Combining Equations \ref{eq:y} and \ref{eq:y1} yields
\begin{equation}
\begin{aligned}
&\int \theta(\vec{q}) \partial_t  f(\vec{q},t) d\vec{q} \\
&=  \alpha \int f(\vec{q},t) \sum_{j \in \{1,2\}} p_j(\vec{q}) [\vec{e}_j \cdot  ( \vec{r}_t - \vec{q})]  \partial_{q_j}\theta(\vec{q}) d\vec{q} .
\end{aligned}
\end{equation}
Using integration by parts, for a typical PDF such that $f(\vec{q},t)$ approaches $0$ as $q_1,q_2 \to \pm \infty$, we have 
\begin{align}
 & \int \theta(\vec{q}) \partial_t  f(\vec{q},t)  d\vec{q}   \notag \\
&\quad = - \alpha \int   \theta(\vec{q})  \sum_{j \in \{1,2\}}  \partial_{q_j}  \left[ f(\vec{q},t)   p_j(\vec{q}) \left [\vec{e}_j \cdot  ( \vec{r}_t - \vec{q} ) \right ] \right ] d\vec{q}. 
\end{align}
Note that this equation holds for any  test function $\theta(\vec{q})$. 
From this, we obtain 
our key result --- the \emph{continuity equation model} (CEM) --- as follows:
\begin{equation}
\begin{aligned}
     \partial_t  f(\vec{q},t)  +  \alpha \! \sum_{j \in \{1,2\}} & \partial_{q_j} \! \left[ f(\vec{q},t)   p_j(\vec{q}) ( R(a_j,\vec{o}_t,t) - q_{j}) \right] = 0 
    \\ 
     & \text{s.t.} \quad o_{j,t} = \mathbb{E}[p_j(\vec{Q}_t)]
\end{aligned}
\label{eq:pde}
\end{equation}
where $p_j$ is the probability of using action $a_j$ given by the Q-values and the applied exploration mechanism, and $R(a_j,\vec{o}_t,t)$ is the reward function of the population game. This equation describes the temporal evolution of the PDF $f(\vec{q},t)$. Recall that the value of $f(\vec{q},t)$ at a given point $\vec{q}$ is asymptotically the fraction of agents having their  Q-values equal to $\vec{q}$ in the population at time $t$. Thus, this equation expresses over time how a system of Q-learners concentrates on each possible pair of Q-values $\vec{q}$ in the Q-value space $\mathbb{R}^2$ during their repeated plays of population games.

\subsection{Key Properties of the Model}
\paragraph{Formal connections to continuity equations.}
We recognize that Equation \ref{eq:pde} can be viewed as a \textit{continuity equation} well known in physics.
To see this, let $\rho= f(\vec{q},t)$ and $\vec{v} = [v_1, v_2]^\top$ such that  $v_j = \alpha p_j(\vec{q}) \left ( R(a_j,\vec{o}_t,t) - q_{j} \right ), \forall j \in \{1,2\}$, and we recover the continuity equation 
\begin{equation}
\label{eq:continuity}
    \partial_t \rho + \sum_{j\in \{1,2\}} \partial_{q_j} (\rho v_j) = 0.
\end{equation}
A continuity equation describes the transport of some quantity, such as mass and energy, in a physical system. 
In the equation, $\rho$ is the density of the quantity and $\vec{v}$ is the velocity field for that quantity.
Thus, this suggests a physical interpretation of our model --- the Q-learning dynamics in population games is analogously the transport of the agent mass in the $m$-dimensional Q-value space (where $m$ is the number of actions) such that the velocity of the agent mass is given by $\alpha p_j(\vec{q}) (R(a_j,\vec{o}_t,t) - q_{j})$ in each direction.
The essence of the continuity equation is a local form of conservation law, indicating that over time, the agent mass can neither be created nor destroyed. In addition, the agent mass moves in a continuous flow and cannot ``teleport'' from one position in the Q-value space to another.

\paragraph{Applicability to different exploration mechanisms.}  Balancing the exploitation-exploration trade-off is challenging in MAL and it has been shown that the choice of exploration mechanism has a significant impact. 
Our model wraps the probability of taking each action into a term $p_j(\vec{q})$ in Equation~\ref{eq:pde}.
As such, the CEM can describe the dynamics of Q-learning with different exploration mechanisms simply by instantiating $p_j(\vec{q})$. 
Due to space constraints, we illustrate this in Section 4 of the supplementary by contrasting the Boltzmann exploration and the power probability form~\citep{camerer1999experience} 
common in behavioral economics.

\begin{table*}[tb]
    \centering
    \small
\begin{tabular}{cccc}
\toprule
Game & Available Actions & Reward Function & Remarks \\ \hline
 \multirow{2}{*}{Public Goods}  & Cooperate ($a_1$) & $1.5\times o_{1,t} -0.5$ & unique NE:  \\
 & Defect ($a_2$) & $ 1.5 \times o_{1,t} $  & all  defects\\ \hline
\multirow{2}{*}{Product Choice}& Mac ($a_1$) & $0.5 + o_{1,t}$ & two pure-strategy NE: \\
 & Windows ($a_2$) & $ 1 - 1.5 \times o_{1,t}$ &  all choose Mac or all choose Windows \\ \hline
  \multirow{2}{*}{Time-Varying Product Choice}& Mac ($a_1$) & $0.5 + o_{1,t}$  & two pure-strategy NE:  \\ 
  & Windows ($a_2$) & $1 - 1.5 \times o_{1,t} \text{ if } t\in [0,10], \,  1.5 - o_{1,t} \text{ else}$ & all choose Mac or all choose Windows  \\ \hline
   \multirow{2}{*}{El Farol Bar}& Stay Home  ($a_1$) & $0$  & numerous pure-strategy NE: \\
 &  Go to the Bar ($a_2$) & $1 \text{ if } o_{2,t} \in [0, 0.6), \, -1 \text{ else}$ & exactly 60\% of agents go to the bar \\ 
 \bottomrule
    \end{tabular}
    \caption{\small Summary of the game configurations considered in this work. NE is short for Nash equilibrium. $\vec{o}_t=[o_{1,t}, o_{2,t}]^\top$ is the population state at time $t$ where $o_{1,t}$ and $o_{2,t}$ represent the proportions of agents that use actions $a_1$ and $a_2$, respectively, at time $t$. }
        \vspace{-0.3in}
    \label{tab:my_label}
\end{table*}

\paragraph{Relations to mean dynamics.}
Equation~\ref{eq:pde} can also be used to investigate the dynamics of the population state $\vec{o}_t$. The dynamics of  $\vec{o}_t$  corresponds to the conventional definition of \emph{mean dynamics} in evolutionary game theory~\citep{sandholm2015population}. 
For each action $a_j \in \mathcal{A}$, the time derivative of  $o_{j,t}$ (as derived in the supplementary) 
is 
\begin{equation}
\label{eq:dext}
\begin{aligned}
  \frac{d o_{j,t}}{d t} 
  = \alpha  \sum_{j \in \{1,2\}} \int   f(\vec{q},t)   p_j(\vec{q}) (R(a_j,\vec{o}_t,t)- q_j) \partial_{q_j} p_j(\vec{q})  d\vec{q}. 
\end{aligned}
\end{equation}
Similarly, we also have the dynamics of the mean Q-value 
\begin{align}
\label{eq:deqt}
    \frac{d \mathbb{E}[Q_{j,t}]}{d t} 
    = \alpha  \int   f(\vec{q},t)   p_j(\vec{q}) \left( R(a_j,\vec{o}_t,t)- q_j \right)  d\vec{q}.
\end{align}
Note that the typical approach~\cite{sandholm2015population} of analyzing the mean dynamics in evolutionary game theory \textit{cannot} be applied here because the mean dynamics has an explicit dependence on the PDF $f(\vec{q},t)$.

\paragraph{Reducibility to ODEs for homogeneous populations.}
Consider a homogeneous population where agents have the same initial Q-values $\vec{Q}_0^{\ast}$, i.e. $\vec{Q}_{0}^{i} = \vec{Q}_0^{\ast},  \forall i\in \mathcal{N}$.
The PDF $f(\vec{q},t)$ at time $t=0$ can be represented by a Dirac delta function, i.e. $f_0= \delta(\vec{Q}_0^{\ast})$.  
According to Equation~\ref{eq:pde}, 
the probability density will remain concentrated on a single point at the next time step and beyond.
Let $Q^{\ast}_{j,t}$ denote the Q-value of action $a_j$ (the same for every agent) at time $t$. 
Based on Equation~\ref{eq:deqt}, we obtain the dynamics of the Q-value
\begin{equation}
\label{eq:deqthomo}
\begin{aligned}
  \frac{d Q^{\ast}_{j,t}}{dt} & =     \frac{d \mathbb{E}[Q_{j,t}]}{d t} 
   = \alpha p_j(\vec{Q}^{\ast}_t) ( R(a_j,\vec{o}_t,t) - Q^{\ast}_{j,t})
\end{aligned}
\end{equation}
where $o_{j,t} = p_j(\vec{Q}_t^{\ast})$.
We observe that there is no explicit dependence on the PDF $f(\vec{q},t)$. This suggests that for homogeneous populations, the dynamics of Q-learning in population games can be characterized by a system of coupled ODEs (Equation~\ref{eq:deqthomo} for each action $a_j \in \mathcal{A}$).

\vspace{-0.15in}
\paragraph{Extension to 2-player \& n-player games.}
A homogeneous agent population can be viewed as an individual agent, since every agent in the homogeneous population has the same initial Q-values and develops the same Q-values afterwards.  Hence,  Equation~\ref{eq:deqthomo} can be extended to Q-learning in 2-player or n-player games. 
Consider a \emph{finite} set $\mathcal{M}$ of agents such that $|\mathcal{M}|=2$ for 2-player games and $|\mathcal{M}|=n$  for n-player games. 
To achieve this, for each pair of agent $i \in \mathcal{M}$ and action $a_j \in \mathcal{A}_i$ (where $\mathcal{A}_i$ is the set of actions available to agent $i$), one can maintain a separate ODE
\begin{equation}
\label{eq:deqtfinite}
  \frac{d Q^{i}_{j,t}}{dt}  = \alpha p_j(\vec{Q}^{i}_t) ( r_{j,t}^i - Q^{i}_{j,t})
\end{equation}
where $r_{j,t}^{i}$ is the reward of agent $i$ that takes action $a_j$ in the 2-player or n-player games at time $t$.
Compared with Equation~\ref{eq:deqthomo}, the main difference is the substitution of reward functions.
Note that the dynamics described by this equation and the REM (Equation~\ref{eq:replicator}) are not equivalent due to the reasons explained in Section 4.2.
We leave the study of Q-learning dynamics in 2-player and n-player games using Equation~\ref{eq:deqtfinite} to future work.

\paragraph{Generalization of the 2-player FPE model.} The continuity equation model is a generalization of the Fokker-Planck equation model~\cite{hu2019modelling} beyond 2-player games and Boltzmann exploration. The setting in \cite{hu2019modelling} considers agents that each play a 2-player symmetric game with \emph{every} other agent in a large population.
Our key observation is that for each 2-player symmetric game in their setting, there exists a reward-equivalent population game.
Let $U(a_1, a_2)$ be the reward function of a 2-player-2-action symmetric game. For  any agent $i$ in the population $\mathcal{N}$, at time $t$, its reward for using action $a_1$ in the 2-player games with every other agent $j$ in the population is 
\begin{equation}
\label{eq:reward}
\begin{aligned}
    \sum_{j\in \mathcal{N}\setminus \{i\}} \sum_{k \in \{ 1,2 \} }  \mathbbm{1}(\text{$j$ uses $a_k$}) U(a_1,a_k) 
    = \frac{1}{n-1} \sum_{k \in \{ 1,2 \} }  U(a_1,a_k) o_{k,t}
\end{aligned}
\end{equation}
where $n = |\mathcal{N}|$ and $\mathbbm{1}(\cdot)$ is the indicator function.
The left hand side is the agent's reward in the 2-player games, and the right hand side corresponds to the reward-equivalent population game.
Therefore, our model can describe the learning dynamics of each 2-player symmetric game in the setting of~\cite{hu2019modelling}.

\section{Experiments}
\begin{figure*}[!tb]
        \centering
    \includegraphics[width=0.90\linewidth]{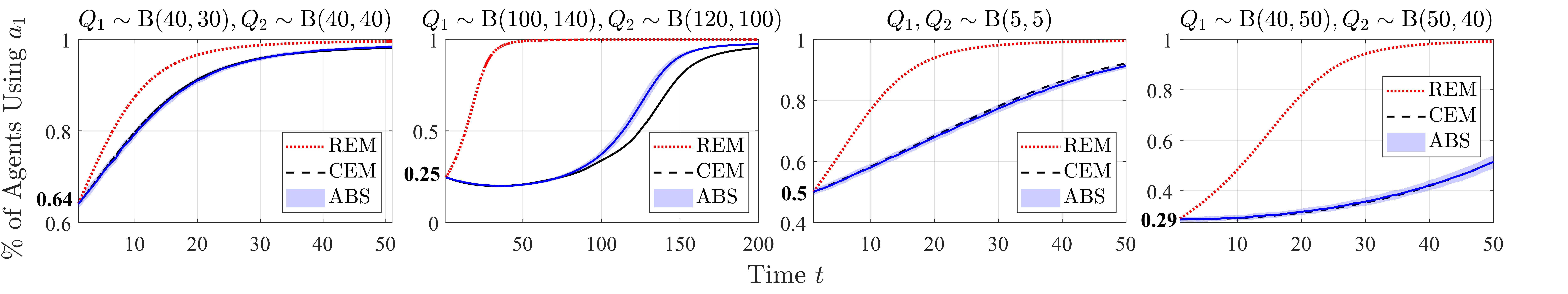}
    \vspace{-0.1in}
    \caption{\small Comparison of the population share $o_{1,t}$ of playing action $a_1$ in the product choice game for heterogeneous MASs.
    $\mathrm{B}$ denotes the Beta distribution with support $[-1.5, 1.5]$.
    Our CEM better captures the qualitative and quantitative dynamics of populations across  different  initial Q-value distributions.
    In particular, as shown in the second plot, our CEM captures a somewhat surprising phenomena  --- although the critical mass for action $a_1$ was reached initially, the population did not behave exactly the same as predicted by the network effect; rather, the population exhibited a decreasing trend in the use of action $a_1$ for the first 50 time steps.
    }
    \vspace{-0.15in}
    \label{fig:exphetero}
\end{figure*}
In this section, we first validate that our CEM indeed provides a more accurate description of the learning dynamics in population games, compared with the REM.
Then, through a concrete example, we show that our CEM can provide insights on the effects of algorithm parameters, which potentially guides parameter tuning.

\subsection{Manifesting Nontrivial Temporal Dynamics}
For MAL, 
the learning dynamics can be far from trivial even in a simple 2-player matrix game~\citep{boone2019darwin,nagarajan2020chaos,sato2002chaos}.
However, with an accurate formal model, the temporal evolution of a MAS manifests itself.
To validate the descriptive power of our CEM, we considered three typical types of population games: product choice games, public goods games, and the El Farol bar problems.
The game configurations are summarized in Table 1.
Due to space restrictions, here we focus on the product choice games. The results of the other games are presented in Section 3 of the supplementary. 
Unless otherwise stated, the Boltzmann temperature is $3$ and the learning rate is $0.05$.

The product choice games model the \emph{network effect} phenomena commonly observed in economics. When the network effect is present, the value of a product monotonically increases in the number of its users; however, for the network effect to take hold, the number of users needs to reach a critical mass. Here, the critical mass for action $a_1$ is $20\%$ of agents in the population.
We compared our CEM and the REM in terms of which model provides a more accurate description of the population state $\vec{o}_t$ over time, given different settings of initial Q-value distributions.

Figure~\ref{fig:exphetero} clearly shows that our CEM indeed has better descriptive power across all the considered settings.
In particular, for the second setting in which  around $25\%$ of agents takes action $a_1$ at time $t=0$,
the network effect phenomena suggests that the population share of action $a_1$ should increase  since the critical mass for action $a_1$ has been reached.
However,
our CEM accurately captures a somewhat surprising phenomena --- although the critical mass for action $a_1$ was reached at time $t=0$, the population share of $a_1$ first experienced a decreasing trend for around 50 time steps and gradually increased thereafter.
Such interesting phenomena, unfortunately, was not captured by the REM, which predicts that the population quickly flocks to the use of action $a_1$.

\vspace{-0.1in}
\subsection{Shedding Light on Algorithm Parameters}
Inspired by Leonardo et al.'s  work~\cite{leonardos2020catastrophe,leonardos2021exploration}, we utilize our CEM to investigate the exploration parameter (the Boltzmann temperature $\tau$), which balances the  exploitation-exploration trade-off in Q-learning. 
Traditionally, finding an appropriate  exploration parameter is cumbersome and requires many simulation runs.
With our CEM, the effects of the parameter on the long-term learning behavior are readily observable and thus, one can draw  insights into appropriate choices of the parameter for a desired learning behavior.

\begin{figure*}[htbp]
\centering
\begin{minipage}[t]{0.73\textwidth}
\centering
\includegraphics[width=0.92\linewidth]{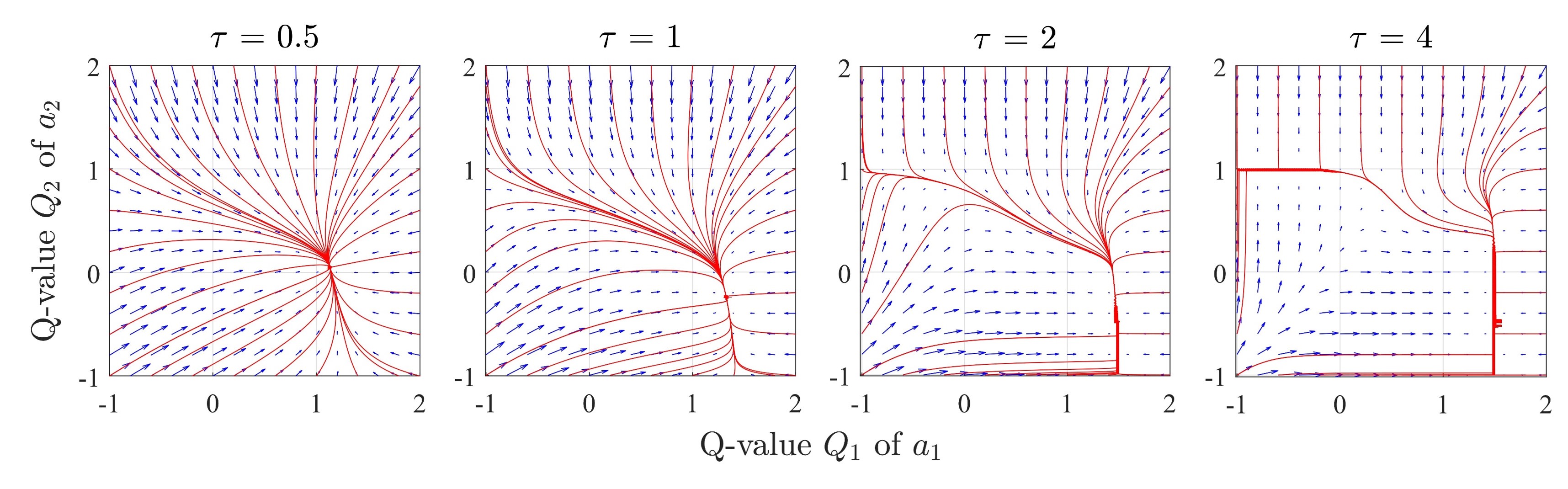}
\vspace{-0.14in}
\caption{\small Slope field plots of the Q-value dynamics in homogeneous MASs that play the product choice game. The arrows give the slope ${d Q_2}/{d Q_1}$, the red lines highlight the trajectories, and the points where the red lines converge are the fixed points ($dQ_2/dQ_1 = 0$). Our CEM shows how the Boltzmann temperature $\tau$ affects the position of the fixed point and the manner by which it is reached. In particular, our CEM suggests that a high temperature $\tau$ causes the phenomena shown in the second plot of Figure 2. }
    \vspace{-0.1in}
\label{fig:tauhomo}
\end{minipage}
\hfill
\begin{minipage}[t]{0.24\textwidth}
\centering
\includegraphics[width=0.92\textwidth]{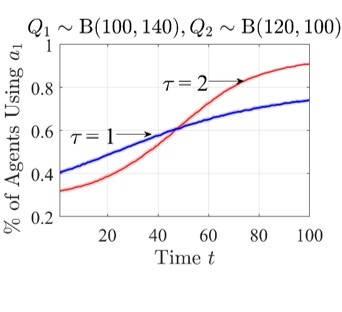}
\vspace{-0.15in}
\caption{\small Simulation results verify CEM's  prediction that  decreasing the temperature $\tau$ makes the phenomena shown in the second plot of Figure 2 disappear. }
\label{fig:tauhetero}
\end{minipage}
\end{figure*}

As a concrete example, we consider homogeneous populations that play the product choice game.
For homogeneous populations, our CEM can be reduced to coupled ODEs (Equation~\ref{eq:deqthomo}).
Figure~\ref{fig:tauhomo} visualizes the solution to these equations, given different choices of temperature $\tau$.
The plotted slope fields illustrate the long-term learning process ---  a homogeneous population starting with a given pair of Q-values (i.e., at a given point in the field) will adapt its Q-values following the trajectory specified by the field. 
Using the slope fields obtained by our CEM, we can readily observe the effects of $\tau$ on the fixed point (or steady state) to which a population converges and the manner by which the fixed point is reached.

Let us consider the populations that start with negative Q-values of two actions (i.e. the left bottom corner of the plots). 
We observe that in general, as $\tau$ increases, the fixed point moves towards the direction of a higher Q-value ($Q_1$) of action $a_1$ and a lower Q-value ($Q_2$) of action $a_2$, suggesting that the population will stabilize with a larger population share of action $a_1$.
When $\tau$ is not larger than $1$, the Q-values of two actions increase almost linearly to reach the fixed point.
However, when $\tau$ becomes larger, the ways by which a population reaches the fixed point change drastically. 
For the population that starts with a higher $Q_1$ (i.e. below the diagonal), $Q_1$ surges directly to reach the fixed point. 
In contrast, for populations that start with a higher $Q_2$ (i.e. above the diagonal),  $Q_2$ initially surges but  gradually decreases to the fixed point.

The above observations lead to the following insights on the choice of Boltzmann temperature $\tau$ in the product choice games:
(i) a higher temperature should in general lead to more agents eventually using action $a_1$, (ii) with a sufficiently low temperature (e.g. $\tau\leq 1$), the Q-values and the policy of agents should quickly become stable, and (iii) with a sufficiently high temperature (e.g. $\tau > 1$), the populations that start with a higher Q-value of action $a_2$ stick to using  $a_2$ for a significant period of time before finally converging to use action $a_1$. 

We find that these insights not only directly apply to homogeneous populations, but also potentially guide parameter tuning for the more general heterogeneous populations.
In particular, the last insight suggests a cause of the phenomena that we observed in the second plot of Figure~\ref{fig:exphetero}: the high Boltzmann temperature.
To validate this, we decreased the temperature parameter in agent-based simulations. 
As shown in Figure~\ref{fig:tauhetero},   given
the same initial Q-value distribution and learning rate as in the second plot of Figure~\ref{fig:exphetero},  with a lower temperature,
the population share of action $a_1$ increases over time and the phenomena of interest disappears. 

\section{Discussion}
In this paper, we examined the dynamics of Q-learning in population games. We began by pointing out the limitations of the replicator equation model when applied to this setting. As a remedy, we developed our continuity equation model (CEM) and analyzed its key properties.
We provided extensive numerical validation for the descriptive power of our model and also illustrated two use cases.

In general, our model works well for a sufficiently large agent population (e.g. consisting of at least hundreds of agents) with a continuously differentiable probability density function (PDF) of Q-values. However, our model may be inaccurate when (i) the agent population is small (e.g. consisting of only dozens of agents), and (ii) the PDF of Q-values is not smooth. Nevertheless, we believe our model is widely applicable; the assumption of a large agent population is standard as population games (by default) are frameworks for large-scale MASs. Regarding the PDF of Q-values, many common probability distributions (e.g. Beta and normal) enjoy the smoothness property. 

We believe that our CEM is an important step towards more general models; as future work, it would be interesting to consider stateful population games, variants of Q-learning, and population games with multiple populations (e.g., with different kinds of agents including  human models~\cite{Teh21}) or network structure. 
We hope that our work can encourage more work along this line of research.

\vspace{-0.1in}
\section*{Acknowledgments}
This research is supported by the National Research Foundation Singapore under its AI Singapore Programme (Award Number: AISG-RP-2019-011).
The work described in this paper was partially supported by a grant from the Research Grants Council of the Hong Kong Special Administrative Region, China (Project No. CUHK 14209321).

\balance
\bibliographystyle{ACM-Reference-Format}
\bibliography{ref}

\end{document}